\documentclass[12pt]{iopart}

\begin{document}

\title{Bell inequalities for random fields}
\date{$23^\mathrm{rd}$ May, 2006.}
\author{Peter Morgan}
\address{Physics Department, Yale University, CT 06520.}
\ead{peter.w.morgan@yale.edu}

\begin{abstract}
The assumptions required for the derivation of Bell inequalities are not
satisfied for random field models in which there are \emph{any} thermal
or quantum fluctuations, in contrast to the general satisfaction of the assumptions
for classical two point particle models.
Classical random field models that explicitly include the effects of quantum
fluctuations on measurement are possible for experiments that violate Bell inequalities.
\newline
\textit{J. Phys.} \textbf{A}: \textit{Math. Gen.} \textbf{39} (2006) 7441-7455.\newline
http://stacks.iop.org/0305-4470/39/7441\newline
doi: 10.1088/0305-4470/39/23/018
\end{abstract}

%\keywords{Bell inequalities; random fields; quantum fields}
\pacs{03.65.Ud, 03.70.+k, 05.40.-a}
\maketitle

\newcommand\Intd{{\mathrm{d}}}
\newcommand{\kT}{{{\mathsf k}_B T}}
\newcommand{\Half}{{\frac{1}{2}}}
\newcommand{\RA}{\mathcal{R}_A}
\newcommand{\RB}{\mathcal{R}_B}
\newcommand{\PastRA}{\textsl{Past}(\RA)}
\newcommand{\PastRB}{\textsl{Past}(\RB)}
\newcommand{\RX}{\mathcal{R}_X}
\newcommand{\RY}{\mathcal{R}_Y}
\newcommand{\PastRX}{\textsl{Past}(\RX)}
\newcommand{\PastRY}{\textsl{Past}(\RY)}
\newcommand{\Note}[1]{{\textbf{[[}\textsl{#1}\textbf{]]}}}
\newcommand{\VSbefore}{\vspace{-0.0cm}}
\newcommand{\VSafter}{\vspace{-0.0cm}}

\section{Introduction}
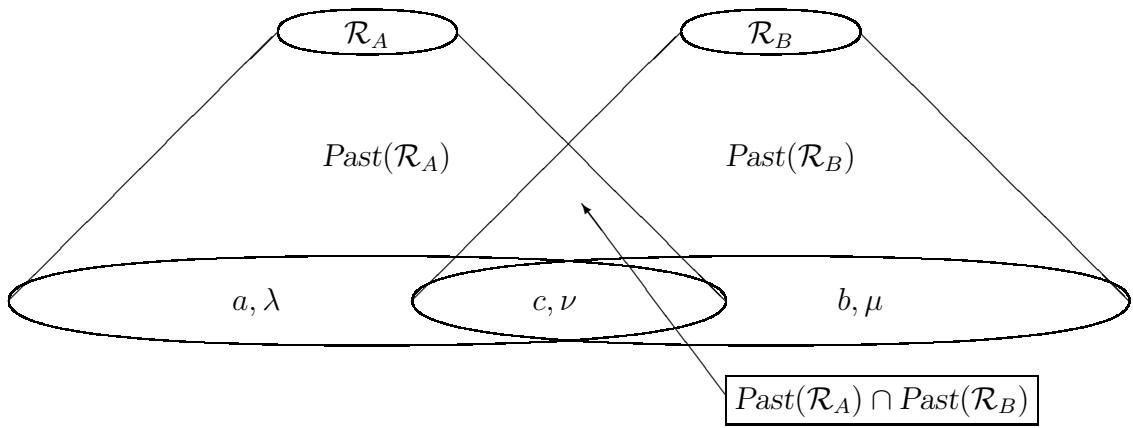
\begin{figure}[htb]
\setlength{\unitlength}{0.038\columnwidth}
\begin{picture}(26,9) 
\multiput(0,2.4)(9,0){2}{\line(1,1){6}}
\multiput(10,8.4)(9,0){2}{\line(1,-1){6}}
\multiput(0,2.4)(9,0){2}{\qbezier(0,0)(0,1)(8,1)}
\multiput(0,2.4)(9,0){2}{\qbezier(8,1)(16,1)(16,0)}
\multiput(0,2.4)(9,0){2}{\qbezier(0,0)(0,-1)(8,-1)}
\multiput(0,2.4)(9,0){2}{\qbezier(8,-1)(16,-1)(16,0)}
\multiput(6,8.4)(9,0){2}{\qbezier(0,0)(0,.5)(2,.5)}
\multiput(6,8.4)(9,0){2}{\qbezier(2,.5)(4,.5)(4,0)}
\multiput(6,8.4)(9,0){2}{\qbezier(0,0)(0,-.5)(2,-.5)}
\multiput(6,8.4)(9,0){2}{\qbezier(2,-.5)(4,-.5)(4,0)}
\put(7.5,8.15){$\RA$}
\put(16.5,8.15){$\RB$}
\put(7,5.4){$\PastRA$}
\put(16,5.4){$\PastRB$}
\put(16,0){\framebox{$\PastRA\cap\PastRB$}}
\put(16,.3){\vector(-3,4){3.2}}
\put(5,2.2){$a,\lambda$}
\put(18.5,2.2){$b,\mu$}
\put(11.7,2.2){$c,\nu$}
\end{picture}
\caption{\label{Fig1}The association of random variables to space-time regions.}
\end{figure}

Bell \cite[Chap. 7, originally 1976]{Bell} shows that from a definition
of local causality, we can derive Bell inequalities for observable statistics
associated with two space-like separated regions $\RA$ and $\RB$
(see figure \ref{Fig1}), and that quantum theory does not satisfy the same
inequalities.
Bell's derivation uses the language of ``beables'', but the mathematics requires
only that random variables are associated with regions of space-time, prompting
us here to introduce \textit{random fields} to remove the niceties of Bell's concept of
``beables''.
Bell introduces a distinction between fields that are ``really supposed to \emph{be} there'',
the ``beable'' fields, and fields that are not real, which is an ontological distinction
that is downplayed here (see \ref{BeablesDescription} for a brief account of ``beables'',
including Bell's examples of fields that are and are not ``beables'').

Almost all experiments that investigate physics at a small scale require the collection of
statistics; it is barely possible otherwise to identify any regularities.
The mathematical tool that we use as an idealization of statistics is the random variable.
At its most general, any indexed set of random variables is a random field, but in the context
of physical descriptions that are placed in space-time, the simplest index set is a lattice
of points (see, for example, Vanmarcke \cite{Vanmarcke}).
This paper will focus, however, on \textit{continuous} random fields (see \ref{RandomFields}
and, for example, Rozanov \cite{Rozanov}), for which the index set is the Schwartz space of
functions on Minkowski space.
A continuous random field can be understood as a random variable-valued distribution.
We will focus on continuous random fields because they are very closely parallel to the operator-valued distributions of quantum field theory (see, for example, Haag \cite[Chap. II]{Haag}).

Against Bell \cite[Chap. 7]{Bell}, Shimony, Horne, and Clauser \cite[originally 1976]{SHC}
show that if random variables associated with $\PastRA - \PastRB$ and with
$\PastRB - \PastRA$ are correlated with random variables associated with
$\PastRA \cap \PastRB$, then a model need not satisfy the Bell inequalities.
Bell \cite[Chap. 12, originally 1977]{Bell} admits this, but finds that random 
variables associated with $\PastRA \cap \PastRB$ have to be correlated with
instrument settings in $\RA$ and in $\RB$.
Arguing that such a requirement is unreasonable, Bell calls it a
``conspiracy'' \cite[Chap. 12, p. 103]{Bell}.
Bell's argument and Shimony, Horne, and Clauser's comments are brought
together in a review article by d'Espagnat \cite{dEspagnat}.

The violation of Bell inequalities by experiment has imposed a moratorium on the construction
of classical models, because it is generally acknowledged that the assumptions required to
derive Bell inequalities are satisfied for classical two point particle models (and the same is
urged here).
However, although Bell's argument is quite clear-cut for classical particles and for any
classical systems that are separated from measurement devices in a well-defined way, the
assumptions required to derive Bell inequalities are not usually satisfied for
random field models if there are \emph{any} thermal or quantum fluctuations\footnote{A clear
distinction between quantum and thermal fluctuations, at least for free fields, is made
in \cite{Morgan}, and is briefly described in \ref{RandomFields}.}.
The correlation that is called ``conspiracy'', more than just being natural, is always
present for random fields if there are any thermal or quantum fluctuations\footnote{For random
fields at thermal equilibrium, 2-point correlations at space-like separation are generally non-zero.
At thermal equilibrium, correlations decay more-or-less exponentially with increasing space-like
separation; for the trivial Gaussian model in three dimensions, for example, the 2-point
connected correlation function is proportional to $e^{-m x}/x$ as a function of space-like
separation $x$ \cite[\S 8.1]{BinneyEtAl}.
These thermal equilibrium correlations at space-like separation already do not satisfy
the assumptions required to derive Bell inequalities.
We can also construct non-equilibrium states of the Gaussian model in three dimensions
in which correlations decay more-or-less like the thermal equilibrium state outside a bounded
space-time region $\mathcal{E}$ that contains an experiment, but with arbitrary correlations
within $\mathcal{E}$, just because classically we have free control of initial conditions,
still more violating the assumptions.},
even though it is not at all natural for classical two-point particle models.
Even if this were all, we would not be able to derive Bell inequalities for random
fields, but section \ref{Derivation} shows that there are numerous other correlations,
\emph{all} of which must also be assumed to be \emph{identically} zero, whereas all
of them are nontrivial for random fields if there are any thermal or quantum
fluctuations.

The literature on Bell inequalities for ``beables'' is quite sparse, and has
not changed the general perception that Bell \cite[Chap. 12]{Bell} more-or-less
closes the discussion.
The more general literature on Bell inequalities, for which the assumptions
required to derive Bell inequalities are discussed quite clearly by
Valdenebro \cite{Valdenebro}, has come to the same conclusion.
\ref{TraditionalBell} discusses the relationship between Bell
inequalities for ``beables'' and the more general literature, in the light of
Section \ref{Derivation}.
In a later paper, Bell \cite[Chap. 16, Originally 1981]{Bell} claims to address the
question of fields,
\begin{quote}
`Finally you might suspect that the very notion of a particle, and particle orbit, ...,
has somehow led us astray. Indeed did not Einstein think that fields rather than particles
are at the bottom of everything? So the following argument will not mention particles,
nor indeed fields, nor any other particular picture of what goes on at the microscopic
level',
\end{quote}
but the argument he then makes is almost exactly the same argument as is made in the papers
referred to above and described in detail below.
It is perhaps just because of the abstraction of his argument, with all mention of particles
or fields removed, that he does not identify the very different nature of the necessary
assumptions that numerous correlations must be precisely zero for the substantially different
cases of two point particles, of $\mathrm{C}^\infty$ fields, and of random fields.

The fundamental definition required for the derivation of Bell inequalities when
random variables associated with space-time regions are the focus of our attention
is of \emph{local causality}.
The two competing definitions given by Bell and by Shimony, Horne, and Clauser are
described in section \ref{Definitions}.
Section \ref{Derivation} will reproduce Bell's mathematical argument in the
form given by d'Espagnat \cite{dEspagnat} to allow the assumptions required for the
derivation to be highlighted.
Sections \ref{ConspiracySection} and \ref{CorrelationSection} discuss in detail the
various correlations that have to be assumed to be precisely zero, then
section \ref{QFTApproach} shows that the violation of Bell inequalities
alone does not justify preferring a quantum field model over a random field model
by considering the similarities between a quantum field theoretic Wigner
quasi-probability description and a random field probability description of
a complete experimental apparatus that violates a Bell inequality.
A quantum field model for a complete experimental apparatus requires as much
``conspiracy'' as a random field model.

The distinction between classical point particles and random fields lies just in
the existence of correlations between random variables associated with
$\PastRA \cap \PastRB$, $\PastRA-\PastRB$, and $\PastRB-\PastRA$, which is not
usual for two classical point particles that emerge from a central source that
lies completely in $\PastRA \cap \PastRB$, but is general for random fields.
The distinction does \emph{not} lie in finite and infinite degrees of freedom ---
the assumptions required to derive Bell inequalities are equally not satisfied for a
sufficiently fine lattice model in which there are thermal or quantum fluctuations.
Because there generally are correlations between random variables associated with
regions at space-like separation in random field models, we may introduce
random field models for complete experimental apparatuses where classical two point
particle models for a measured system are not adequate and where previously only
quantum mechanical models have been thought adequate.

\section{Definitions of local causality}\label{Definitions}
First, the random variables $(a,\lambda)$, $(b,\mu)$, and $(c,\nu)$ are
considered in more detail.
They are associated with the disjoint regions $\PastRA - \PastRB$, $\PastRB - \PastRA$,
and $\PastRA\cap \PastRB$, respectively (see figure \ref{Fig1}).
$a$, $b$, and $c$ are ``non-hidden'' \cite[Chap. 12]{Bell} random variables, instrument
settings that are observed and possibly controlled by the experimenter, while
$\lambda$, $\mu$, and $\nu$ are ``hidden'' random variables, neither observed nor
controlled by the experimenter.
As far as classical physics is concerned, the separation of random variables into
$(a,\lambda)$, $(b,\mu)$, and $(c,\nu)$ is arbitrary, because anything that
is hidden today may be revealed tomorrow and whether we observe or record
random variables makes no difference, so any derivation of Bell inequalities must be
robust under different choices of the separation.
There is nothing about the mathematics of section \ref{Derivation} that will
determine a separation of random variables into $(a,\lambda)$, $(b,\mu)$, and $(c,\nu)$.
The only difference between non-hidden random variables and hidden random variables
will be that we will integrate over all values of hidden random variables and never
integrate over values of non-hidden random variables.
It will be useful to consider three choices in this paper: (1) all of $a$, $b$, 
$c$, $\lambda$, $\mu$, and $\nu$ are non-null sets of random variables;
(2) $\nu$ is a complete set of random variables, so that $c$ is null; and
(3) $c$ is a complete set of random variables, so that $\nu$ is null.

The fundamental definition in Bell's derivation of inequalities is that for a
\emph{locally causal theory}, for $X$ any random variable associated with a
space-time region $\RX$, $X_\cap$ \emph{all} of the random variables associated with
$\PastRX \cap \PastRY$, $X_p$ \emph{some} of the random variables associated with
$\PastRX - \PastRY$, and $Y$ any random variable associated with a space-time region
$\RY$ that is space-like separated from $\RX$, the conditional probability of $X$
given $X_\cap$ and $X_p$ is statistically independent of $Y$,
\begin{equation}\label{LocalCausality}
  p(X|X_\cap, X_p, Y)=p(X|X_\cap, X_p).
\end{equation}\VSafter
(In an abuse of notation, we will write $X$ for an event involving the random variable
$X$; where we discuss several events involving the same random variable,
they will be denoted $X$, $X'$, etc. which may be thought of as shorthand for events
$E_X, E'_X$, etc.
This lets us keep close to the notation of the original papers \cite{Bell,SHC,dEspagnat}.)
This definition of local causality is applied a number of times in Bell's derivation
of inequalities.

Shimony, Horne, and Clauser \cite{SHC}, in contrast, weaken the definition of a
locally causal theory, so that for $X$ and $Y$ as above, but for $X_P$ \emph{all}
of the random variables associated with $\PastRX$, the conditional probability of $X$
given $X_P$ is statistically independent of $Y$,
\begin{equation}\label{LocalCausalitySHC}
  p(X|X_P, Y)=p(X|X_P).
\end{equation}\VSafter
The two definitions of a locally causal theory are the same if $X_p$ happens
to be \emph{all} the random variables in $\PastRX - \PastRY$, but note that
the definition of $X_p$ is so loose that $X_p$ can even be any single random
variable associated with $\PastRX - \PastRY$.
Equation (\ref{LocalCausality}) is presumably supposed to be satisfied for an
\emph{arbitrary} choice of random variables as $X_p$, so it is $X_\cap$ that
is characteristic of equation (\ref{LocalCausality}).
Equation (\ref{LocalCausality}) combines equation (\ref{LocalCausalitySHC}), which
is a much more natural definition of local causality for a classical field theory,
with a principle that correlations only arise because of common causes.
Equation (\ref{LocalCausality}) generalizes an idea that two point particles leave
a central source at the same time as a common cause of two events in regions
$\RA$ and $\RB$, entirely reasonable for a classical two point particle model,
to a much more tendentious idea that there must be a common cause of the two
events even in a random field model for an experiment.

Equation (\ref{LocalCausalitySHC}), however, is not strong enough to allow Bell
inequalities to be derived.
Some of the applications of equation (\ref{LocalCausality}) can be replaced by
applications of equation (\ref{LocalCausalitySHC}), but some cannot.
As well as the well-known ``no-conspiracy'' assumption (discussed in section
\ref{ConspiracySection}), which prohibits correlations between instrument
settings and hidden random variables and is needed whether we adopt equation
(\ref{LocalCausality}) or equation (\ref{LocalCausalitySHC}) as our definition
of a locally causal theory, section \ref{Derivation} further identifies a
``no-correlation'' assumption (discussed in section \ref{CorrelationSection}),
which prohibits correlations between hidden random variables.
The ``no-correlation'' assumption is only needed if we adopt equation
(\ref{LocalCausalitySHC}) as our definition of a locally causal theory.

Although equation (\ref{LocalCausalitySHC}) is a natural definition of local
causality, even it is not satisfied by the signal local and Lorentz invariant
\emph{but analytically nonlocal} dynamics discussed in \cite{Morgan}.
With such a dynamics, $p(X|X_P,Y)\not = p(X|X_P)$, even though the nonlocal
effects of such a dynamics are restricted to heat-equation-like
exponentially reducing tails and signal locality is satisfied.

The derivation of Bell inequalities for random fields also requires an assumption
that apparatus must be independent, which is innocuous if almost all random
variables are hidden, but is not innocuous if almost all random variables are
non-hidden, when significant correlations should be expected.
Section \ref{Derivation} identifies an ``independent-apparatus'' assumption
(discussed in section \ref{ConspiracySection}), and shows the
``independent-apparatus'' assumption to be closely related to the
``no-conspiracy'' assumption.
These three assumptions cannot be considered independently.
Indeed, which assumptions should be considered to be unsatisfied in a model
for an experiment that violates Bell inequalities will depend on what
separation there is of random variables into $(a,\lambda)$, $(b,\mu)$, and $(c,\nu)$,
since the assumptions refer to correlations between non-hidden and non-hidden
random variables (``independent-apparatus''), between non-hidden and hidden random
variables (``no-conspiracy''), and between hidden and hidden random variables
(``no-correlation'').

\section{The derivation of Bell inequalities for random fields}
\label{Derivation}
Assumptions that are required to derive Bell inequalities, and that will be
discussed in sections \ref{ConspiracySection} and \ref{CorrelationSection}, will
be indicated by \Note{Notes in brackets}.
Suppose that $A$ and $B$ are random variables associated with regions $\RA$ and $\RB$.
Recall that the conditional probability $p(X|Y)$ is defined as $p(X|Y)=\frac{p(X,Y)}{p(Y)}$,
so that $p(X,Y|Z)=p(X|Y,Z)p(Y|Z)$.
Applying this first to $p(A,B,\lambda,\mu,\nu|a,b,c)$, we obtain
\begin{equation}
  p(A,B,\lambda,\mu,\nu|a,b,c)=p(A,B|\lambda,\mu,\nu,a,b,c)p(\lambda,\mu,\nu|a,b,c),
\end{equation}
and applying it again to $p(A,B|\lambda,\mu,\nu,a,b,c)$, we obtain
\begin{equation}
  p(A,B|\lambda,\mu,\nu,a,b,c)=p(A|B,\lambda,\mu,\nu,a,b,c)p(B|\lambda,\mu,\nu,a,b,c).
\end{equation}
Applying equation (\ref{LocalCausality}) \emph{or} equation (\ref{LocalCausalitySHC}),
the conditional probability density $p(A|B,\lambda,\mu,\nu,a,b,c)$ is statistically independent
of $b$, $\mu$, and $B$ in a locally causal theory, and similarly
for the conditional probability density $p(B|\lambda,\mu,\nu,a,b,c)$,
\VSbefore
\begin{eqnarray}
   p(A|B,\lambda,\mu,\nu,a,b,c) &=& p(A|a,c,\lambda,\nu),\\
   p(B|\lambda,\mu,\nu,a,b,c) &=& p(B|b,c,\mu,\nu),
\end{eqnarray}
so that 
\begin{equation}
  p(A,B|\lambda,\mu,\nu,a,b,c)=p(A|a,c,\lambda,\nu)p(B|b,c,\mu,\nu).
\end{equation}
Using these, the mean of the product $AB$, given an event $(a,b,c)$, is
\VSbefore
\begin{eqnarray}
M(a,b,c) & = & \int\!\!\!\int\!\!\!\int\sum_{AB}
                    AB\,p(A,B,\lambda,\mu,\nu|a,b,c) d\lambda d\mu d\nu \cr
         & = & \int\!\!\!\int\!\!\!\int\sum_{AB}
                    AB\,p(A,B|\lambda,\mu,\nu,a,b,c)
                    p(\lambda,\mu,\nu|a,b,c) d\lambda d\mu d\nu \cr
         & = & \int\!\!\!\int\!\!\!\int\sum_{AB}
                    AB\,p(A|a,c,\lambda,\nu)p(B|b,c,\mu,\nu)
                    p(\lambda,\mu,\nu|a,b,c) d\lambda d\mu d\nu.\cr&&
\end{eqnarray}
The conditional probability density $p(\lambda,\mu,\nu|a,b,c)$ can also be
rewritten by again applying the definition of conditional probability, as
\VSbefore
\begin{eqnarray}
     p(\lambda,\mu,\nu|a,b,c) & = & p(\lambda,\mu|\nu,a,b,c)p(\nu|a,b,c)\cr
     {} & = & p(\lambda|\mu,\nu,a,b,c)p(\mu|\nu,a,b,c)p(\nu|a,b,c).
\end{eqnarray}
Applying equation (\ref{LocalCausality}), or, through a putative argument
provided by Shimony, Horne, and Clauser \cite{SHC} and discussed in section
\ref{CorrelationSection}, applying equation (\ref{LocalCausalitySHC}), we
can derive
\VSbefore
\begin{eqnarray}
    p(\lambda|\mu,\nu,a,b,c) &=& p(\lambda|\nu,a,b,c), \label{No-Correlation}\\
    p(\lambda|\nu,a,b,c) &=& p(\lambda|\nu,a,c), \label{No-NonlocalA}\\
    p(\mu|\nu,a,b,c) &=& p(\mu|\nu,b,c) \label{No-NonlocalB}
\end{eqnarray}
\Note{$p(\lambda|\mu,\nu,a,b,c) = p(\lambda|\nu,a,b,c)$ is the ``no-correlation''
     assumption; equations (\ref{No-NonlocalA}) and (\ref{No-NonlocalB}) are
     further assumptions, which might be called ``no-nonlocal-conspiracy''
     assumptions, but they will not be directly addressed here},
so the mean of the product $AB$, given the event $(a,b,c)$, is
\begin{equation}\label{MeanProduct}
    M(a,b,c) = \int\overline{A(a,c,\nu)}\>\overline{B(b,c,\nu)}p(\nu|a,b,c)d\nu,
\end{equation}\VSafter
where $\overline{A(a,c,\nu)}$ is the mean of $A$ averaged over the
hidden random variables $\lambda$, given the event $(a,c,\nu)$, and similarly for
$\overline{B(b,c,\nu)}$.

Suppose that $A$ and $B$ satisfy $|A|\leq 1$ and $|B|\leq 1$, so that
$|\overline{A(a,c,\nu)}|\leq 1$ and $|\overline{B(b,c,\nu)}|\leq 1$. If we
also suppose that
\begin{equation}\label{ConspiracyEquation}
   p(\nu|a,b,c) = p(\nu|c)
\end{equation}\VSafter
\Note{$p(\nu|a,b,c) = p(\nu|c)$ is the already known ``no-conspiracy''
assumption}, then we can derive, for distinct events $a$, $a'$, and $b$, $b'$ for
the non-hidden random variables $a$ and $b$,
\VSbefore
\begin{eqnarray}\label{Mdef00}
    |M(a,b,c)\mp M(a,b',c)| & = & \left|\int\overline{A(a,c,\nu)}
      \left[\overline{B(b,c,\nu)}\mp \overline{B(b',c,\nu)}\right]
                  p(\nu|c)d\nu\right| \cr
 \noalign{\vskip 3pt}
    {} & \leq &
    \left|\int\left[\overline{B(b,c,\nu)}\mp \overline{B(b',c,\nu)}\right]
                  p(\nu|c)d\nu\right|,\\
 \noalign{\vskip 6pt}\label{Mdef10}
    |M(a',b,c)\pm M(a',b',c)| & = & \left|\int\overline{A(a',c,\nu)}
      \left[\overline{B(b,c,\nu)}\pm \overline{B(b',c,\nu)}\right]
                  p(\nu|c)d\nu\right| \cr
 \noalign{\vskip 3pt}
    {} & \leq &
    \left|\int\left[\overline{B(b,c,\nu)}\pm \overline{B(b',c,\nu)}\right]
                  p(\nu|c)d\nu\right|,
\end{eqnarray}
\Note{Being able to change $a\rightarrow a'$ without changing $c$ or $b$ and
$b\rightarrow b'$ without changing $c$ or $a$ is the ``independent-apparatus''
assumption, which is insignificant if $a$, $b$, and $c$ are just a few inaccurately
measured variables, but becomes significant if $a$, $b$, and $c$ are extensive or
complete information about the apparatus.}
so that
\begin{equation}\label{BellInequality}
    |M(a,b,c)\mp M(a,b',c)|+|M(a',b,c)\pm M(a',b',c)|\leq 2,
\end{equation}\VSafter
because $\left|\int\overline{B(b,c,\nu)}p(\nu|c)d\nu\right|\le 1$ and
$|\alpha|\le 1 \wedge |\beta|\le 1\Rightarrow|\alpha+\beta|+|\alpha-\beta|\le 2$.
In contrast, for two spin-half particles, we can derive the inequalities
\begin{equation}
    |M(a,b,c)\mp M(a,b',c)|+|M(a',b,c)\pm M(a',b',c)|\leq 2\sqrt{2},
\end{equation}\VSafter
which is essentially the Cirel'son bound \cite{Cirelson}, but, for a random field
model,\newline
\begin{tabular}{l l} 
    {if $p(\lambda|\mu,\nu,a,b,c) \ne p(\lambda|\nu,a,b,c)$,} &
        {--- correlation}\\
    {or $p(\lambda|\nu,a,b,c) \ne p(\lambda|\nu,a,c)$,} &
        {--- nonlocal-conspiracy}\\
    {or $p(\mu|\nu,a,b,c) \ne p(\mu|\nu,b,c)$,} &
        {--- nonlocal-conspiracy}\\
    {or $p(\nu|a,b,c)\ne p(\nu|c)$,} &
        {--- conspiracy}\\
\noalign{\vskip 5pt}
\end{tabular}\newline
then $M(a,b,c)$, $M(a,b',c)$, $M(a',b,c)$, and $M(a',b',c)$ are independent,
satisfying only the trivial inequalities
\begin{equation}\label{MaxInequalities}
  |M(a,b,c)\mp M(a,b',c)|+|M(a',b,c)\pm M(a',b',c)|\leq 4.
\end{equation}\VSafter
Additionally, if we cannot change $a$, $b$, and $c$ independently, we cannot measure
$M(a,b,c)$ \emph{and} $M(a,b',c)$, for example, because we cannot keep $a$ and $c$
\emph{perfectly} unchanged, making both equations (\ref{BellInequality}) and
(\ref{MaxInequalities}) experimentally unrealisable.
All we could measure would be $M(a,b,c)$ and $M(\mathring{a},b',\mathring{c})$, where
$\mathring{a}$ and $\mathring{c}$ are perhaps very close to $a$ and $c$ but not identical.

It is well-known that classical local physics allows the maximum value of 4 to be
saturated \cite{Khalfin,Popescu,Revzen}, but the violation of Bell inequalities has
been considered to require unnatural correlations; this paper argues that the
required correlations are natural for a classical random field.
Classically, quantum mechanics is half-way between the conditions for deriving
Bell inequalities and the maximum violation, when equation (\ref{MaxInequalities})
is satisfied as an equality.
There must, therefore, be principled constraints on initial conditions in a random
field model to ensure the maximum violation is never observed, as well as to allow
some violation.
It is an open problem to find a plausible axiom that restricts random field models
to the same $2\sqrt{2}$ limit as quantum theory, but there are models in
the literature, such as those of Adler\cite{Adler} and 't Hooft\cite{tHooft}, that
are approximately within the general structure of a random field but have generally
been ignored summarily because of an over-zealous belief that violation of the Bell
inequalities rules out all classical models.

\section{The no-conspiracy and independent-apparatus assumptions}
\label{ConspiracySection}
The prohibition of correlations of $a$ with $c$, and of $b$ with $c$,
the ``independent-apparatus'' assumption, is closely related to the
``no-conspiracy'' assumption.
If we suppose that $\nu$ is a complete set of random variables, so that $c$
is null, we can derive in place of equation (\ref{MeanProduct}), supposing
that equations (\ref{No-Correlation}), (\ref{No-NonlocalA}), and
(\ref{No-NonlocalB}) are satisfied,
{
\newcounter{saveeqn}
\setcounter{saveeqn}{\value{equation}}
\renewcommand{\theequation}{$\ref{MeanProduct}^{\nu}$}
\begin{equation}
    M(a,b) = \int\overline{A(a,\nu)}\>\overline{B(b,\nu)}p(\nu|a,b)d\nu,
\end{equation}\VSafter
\setcounter{equation}{\value{saveeqn}}
}
\noindent which requires that
\begin{equation}
    p(\nu|a,b)=p(\nu)
\end{equation}\VSafter
for us to be able to derive Bell inequalities.
If we take $a$ and $b$ to be only instrument settings at the time of the
measurement, with $c$ null, so that $\nu$ is \emph{complete} information about the
\emph{whole} of $\PastRA\cap \PastRB$, the ``no-conspiracy'' assumption asserts that
instrument settings at the time of the measurement must be completely
uncorrelated with the experimental apparatus (which is, after all, almost
entirely in $\PastRA\cap \PastRB$).
Ensuring that instrument settings are \emph{completely} uncorrelated with the
experimental apparatus would seem a remarkable achievement in a random field
theory setting.

Bell argues \cite[Chap. 12]{Bell} that the dynamics of a mechanism to choose the
instrument settings can be made chaotic enough that, even if there are correlations
between $(c,\nu)$ and $(a, b)$, the instrument settings may nonetheless be taken
to be `at least effectively free for the purposes at hand'.
From a classical point of view, this is a remarkable claim.
Either there are correlations in a model for an experiment or there are not.
Correlations that are easy to measure at one time are generally not as easy to
measure at other times, but the practicality of measuring correlations has no
bearing on whether there are correlations, which is in principle unaffected by
whether the evolution is chaotic or not.

In any case, $a$ and $b$ being `free for the purposes at hand' does \emph{not}
imply $p(\nu|a,b)=p(\nu)$. A correlation $p(\nu|a,b)\not=p(\nu)$ does not
``determine'' $a$ and $b$ (or $\nu$), but only describes a statistical relationship
between $a$, $b$, and $\nu$.
If there \emph{is} a correlation between $a$, $b$, and $\nu$, and we arrange or
observe particular statistics for $a$ and $b$, it just must have been the case that
$\nu$ had statistics compatible with the correlation, even though we did not
control or measure $\nu$.
However, $\nu$ is not measured --- by definition, since it's ``hidden'' --- so
we can only surmise whether there is in fact such a correlation and whether the
statistics of $\nu$ are compatible with the correlation.
How measurements are ``determined'' or ``chosen'' is independent of $p(\nu|a,b)$,
because it is only a record of how the unmeasured variables $\nu$ are correlated with
the measurement settings; $\nu$ might entirely ``determine'' or ``choose'' the
measurement settings $a$ and $b$ or not ``determine'' them at all, but have the
same correlations with $a$ and $b$ in both cases.
As Jaynes puts it, rather forcefully, `Bell took it for granted that a conditional
probability $P(X|Y)$ expresses a physical causal influence, exerted by $Y$ on $X$'
\cite{Jaynes}.
 
Bell also argues \cite[Chap. 12]{Bell}
`that the disagreement between locality and quantum mechanics is large --- up
to a factor of $\sqrt{2}$ in a certain sense', and that although the assumptions
identified here are not analytically satisfied, nonetheless they are
``nearly'' satisfied.
First of all, Bell's argument is slightly weakened by the classical limit
being either $2$ or $4$ (Bell omits to mention the latter), depending on
whether we accept \emph{all} the standard assumptions, with $2\sqrt{2}$ as
the intermediate quantum mechanical limit.
More critically, the standard assumptions discussed here are given as analytic
equalities, which are unable to elaborate Bell's `certain sense'.
A random field model is so general that it is unclear how the no-correlation,
no-nonlocal-conspiracy, no-conspiracy, and independent-apparatus assumptions
could instead be given as physically justifiable limits on inequality (note
that the standard assumptions are problematic just as analytic equalities
between probability distributions, since such a relationship cannot be
supported by experimental statistics, nor, it seems, by analytic argument).

For a random field model to be empirically adequate, there is no requirement
that the assumptions be violated by much, only that the totality of correlations
be such that the dynamical evolution will result in the violation of Bell
inequalities at the time of measurement.
A correlation might be easily measurable at the time of a measurement, but
just because the same correlation is almost always not measurable in practice
at earlier times does not mean it is zero at earlier times.
The chaotic behaviour that Bell invokes to assert that instrument settings
cannot be significantly correlated in fact operates rather against Bell's
overall argument, since then manifest measured correlations between non-hidden
random variables at the time of measurement are all the more likely to correspond
to unmeasurable correlations between hidden random variables before the time of
measurement.

Making slightly different assumptions, suppose that instead of taking $\nu$ to be
complete information, we take $c$ to be complete information, so that $\nu$ is null.
Then we can derive, in place of equation
(\ref{MeanProduct}), again supposing that equations (\ref{No-Correlation}),
(\ref{No-NonlocalA}), and (\ref{No-NonlocalB}) are satisfied,
{
\setcounter{saveeqn}{\value{equation}}
\renewcommand{\theequation}{$\ref{MeanProduct}^{c}$}
\begin{equation}
M(a,b,c)=\overline{A(a,c)}\ \overline{B(b,c)}.
\end{equation}\VSafter
\setcounter{equation}{\value{saveeqn}}
}
\noindent Now to derive equation (\ref{BellInequality}), we have to make only the
``independent-apparatus'' assumption, so that we can change
$a\rightarrow a'$ without changing $c$ (or $b$) and we can change
$b\rightarrow b'$ without changing $c$ (or $a$), with the ``no-conspiracy''
assumption playing no r\^ole.

In quantum field theory, the Reeh-Schlieder theorem \cite{Haag} is typically
thought very awkward, yet the apparatus-dependence it implies is not taken to
rule out quantum field theory.
Recall that as a consequence of the Reeh-Schlieder theorem we cannot change
a quantum field state so that the expected value of a quantum field observable
associated with $\PastRA - \PastRB$ changes without changing the expected
value of almost all quantum field observables associated with both
$\PastRA\cap\PastRB$ and $\PastRB - \PastRA$.
Applied in the context of Bell inequalities for random fields, this is just to
say that it is impossible in quantum field theory to change $a\rightarrow a'$
without changing $c$ at least some of the time, if $c$ is the \emph{complete}
set of observables in $\PastRA\cap\PastRB$.
If instead $\nu$ is the complete set of random variables in $\PastRA\cap\PastRB$,
the Reeh-Schlieder theorem would then be just to say that the ``no-conspiracy''
assumption cannot be satisfied in quantum field theory --- there must be
correlations between $\nu$ and $(a, b)$.

If $c$ is not complete information, the correlations of $c$ with $a$ and $b$
should be expected to lessen as $c$ includes fewer and fewer random variables; the
correlations should not be expected to become identically zero as soon as $c$
is not the complete set of random variables in $\PastRA\cap\PastRB$.
Whether we measure or do not measure random variables in the past should not make any
difference, in a classical model, to whether violation of Bell inequalities
can be observed, but will change the description we give of the correlations
we take to cause the violation.

It is unreasonable to expect the ``independent-apparatus'' and ``no-conspiracy''
assumptions to be satisfied by a random field model when we do not expect them of
quantum field theory --- to do so is to construct a straw man of a theory.
If we insist on a parallel of the Reeh-Schlieder theorem for random fields,
we cannot derive Bell inequalities for random fields.

\section{The no-correlation assumption}
\label{CorrelationSection}
Recall that the ``no-correlation'' assumption, equation (\ref{No-Correlation}),
requires that there are no correlations between the hidden random variables $\lambda$ and
the hidden random variables $\mu$ (that are not screened off by $\nu$, $a$, $b$, and $c$).
There is no empirical way to justify this assumption, simply because it is a
condition imposed on random variables that are \emph{by definition} not measured.
The preference against correlations between instrument settings and hidden random
variables is only tendentiously extensible to justify a prohibition against
correlations between hidden random variables.

Shimony, Horne, and Clauser \cite{SHC} argue that
\begin{quote}
`even though the space-time region in which $\lambda$ is located extends to
negative infinity in time, $\nu$, $a$, $c$ are \emph{all} the beables other
than $\lambda$ itself in the backward light cone of this region, and $\mu$ and
$b$ \emph{do} refer to beables with space-like separation from the $\lambda$
region' (their emphasis)
\end{quote}
to justify deriving equations (\ref{No-Correlation}), (\ref{No-NonlocalA}),
and (\ref{No-NonlocalB}) from equation (\ref{LocalCausalitySHC})
(no additional argument is needed if we take equation (\ref{LocalCausality})
as our definition of local causality).
This argument relies on the unbounded extent of $\PastRA - \PastRB$, so that
on a simple interpretation the only random variables associated with the past
light-cone of $\PastRA - \PastRB$, the region
$\textsl{Past}(\PastRA - \PastRB)$, are $c$ and $\nu$, since $a$ and $\lambda$
are associated with the region $\PastRA - \PastRB$ itself.
Consider, however, that for any time-slice $\PastRA_T$ of $\PastRA$ at time $T$,
we would expect the complete set of random variables associated with $\PastRA_T$
to determine random variables associated with $\RA$ (at least probabilistically),
but we would \emph{not} expect the complete set of random variables associated
with $\PastRA_T\cap\PastRB_T$ to determine random variables associated with $\RA$.
As we consider earlier and earlier time-slices, the contribution from
$\PastRA_T - \PastRB_T$ becomes less and less, but the contribution only
becomes exactly zero in the infinite past if there is \emph{no} incoming
light-like contribution.

Assuming that random variables that determine observables in $\RA$ and $\RB$ must
be associated with the whole of $\PastRA$ and $\PastRB$ goes against the usual
structure of classical physics, which almost always takes initial conditions to
be associated with a time-slice of the past (usually a hypersurface, but at most
a space-time region of finite duration), not to be associated with the whole of
the past.
The competing definitions of local causality, and the whole derivation of
Bell inequalities for random fields, may be put in terms of an arbitrary time-slice
of $\PastRA$ and $\PastRB$.
If we associate the random variables $\lambda$, $\mu$, $\nu$, $a$, $b$, and $c$
with a time-slice of the backward light-cones, not with the whole backward
light-cones, Shimony, Horne, and Clauser's argument fails to justify
deriving equations (\ref{No-Correlation}), (\ref{No-NonlocalA}), and
(\ref{No-NonlocalB}) from equation (\ref{LocalCausalitySHC}).

Shimony, Horne, and Clauser's argument effectively reintroduces common
causation, by requiring that there is no causation associated with the
region outside $\PastRA\cap\PastRB$.
As for the no-conspiracy and independent-apparatus assumptions, there is no
requirement that the no-correlation assumption be violated by much, only
that the totality of correlations of all three kinds be such that the
dynamical evolution will result in the violation of Bell inequalities
at the time of measurement.

Shimony, Horne, and Clauser's argument is contrary to the \emph{intention}
behind equation (\ref{LocalCausalitySHC}), which is that the \emph{complete}
causal past of a region determines its present (again, at least
probabilistically, even if not deterministically).
If we take the trouble to distinguish between equation (\ref{LocalCausality})
and equation (\ref{LocalCausalitySHC}) as definitions of local causality, we
can't argue for equations (\ref{No-Correlation}), (\ref{No-NonlocalA}),
and (\ref{No-NonlocalB}) in a way that quietly negates the distinction.

\section{A quantum field theory approach}
\label{QFTApproach}
We have become used to describing the outcome of Bell violating experiments
using a state in a complex 4-dimensional Hilbert space, in which many detailed
degrees of freedom are integrated out.
If we agree, however, that non-relativistic quantum mechanics is a reduction
of quantum field theory --- as we almost always do --- such a state is a
reduction of a quantum field state in an infinite-dimensional Hilbert space,
which gives the values of quantum field observables associated with the
regions $\RA$ and $\RB$.
If Bell inequalities are violated by observables of a quantum field state, we
would certainly attribute the violation to the experimenters' ingenuity in
ensuring an appropriate initial quantum field state and making appropriate
measurements.
For a quantum field state describing an experimental apparatus that violates
Bell inequalities, the existence of strong correlations between observables at
large space-like separations is a large part of what singles out such states as
special (Bell inequalities are violated slightly even for the vacuum \cite{SW},
but unmeasurably at large space-like separations).
A quantum field state that describes experimental correlations that measurably
violate Bell inequalities at the time of measurement describes correlations in
the remote past different from those of the vacuum state, but, as for a random field
model, differences from the vacuum state may be difficult to detect in the
remote past.
In quasi-probability terms, we have to set up a Wigner quasi-distribution over
phase space in the past that evolves to a Wigner quasi-distribution over phase
space at the time of measurement $t_M$ that violates a Bell inequality in the
regions $\RA$ and $\RB$.

For an equilibrium state of a random field model, correlations between random
variables that violate the assumption of statistical independence at space-like
separation generally decrease more-or-less exponentially fast with increasing
distance, but strong correlations at arbitrarily large distances are possible
for non-equilibrium states.
Indeed, \emph{absolutely any correlations are allowed in a non-equilibrium initial
condition} --- initial conditions of low probability of course require greater free
energy to set up, but we should not forget how difficult it is to construct an
experiment that violates Bell inequalities at large space-like separations.
In a random field model, we have to set up a probability distribution
over phase space in the past that evolves to a probability distribution over phase
space at time $t_M$ that violates a Bell inequality in the regions $\RA$ and $\RB$,
but this is no greater ``conspiracy'' than is apparent in the full quantum field
state for the experiment (to be explicit, note the parallel between the Wigner
quasi-distribution description and the classical probability description).

The statistics we observe for random variables in the region $\RA\cup\RB$ are no more
than classical initial conditions. We cannot rule out any classical dynamics,
whether local or non-local, without Bell's other assumptions, which amount to a
claim that unobserved initial conditions at earlier times cannot, for
\textit{a priori} reasons, be correlated in such a way that the observed
initial conditions at the time of measurement are as we observe them.
There are often significant reasons for preferring a quantum field model over
a random field model, such as ease of computation, years of familiarity, and the
analytical power of the mathematics of Hilbert spaces, but the violation of Bell
inequalities is not conclusive.

The correlations we have discussed here commit us to very little, if we take an
equally empiricist approach to random fields as we take to quantum fields:
correlations just exist; we do not have to assume that they are caused by
common (or any other kind of) causes.
Classical physics has generally taken initial conditions to be more-or-less explained
by earlier initial conditions, with no final explanation being essential.

\section{Discussion}
We have described the previously identified difference between Bell's definition of
a locally causal theory, which insists that correlations have to be the result of
common causes, and Shimony, Horne, and Clauser's definition, which does not.
The assumption that there \emph{is} a common cause for separated events is an
\textit{a priori} constraint on unmeasured initial conditions at earlier times.
This is quite a natural assumption for a classical two particle model, because the
two particles are imagined to be emitted from a single point in the past, but it is
a strong and unjustifiable assumption for a random field model.
We have described numerous assumptions that random variables may not be correlated
with other random variables, all of which are necessary for Bell inequalities
to be derived, but none of which are generally satisfied for a random field in
the presence of thermal or quantum fluctuations.

We have also seen that the opprobrium of ``conspiracy'' as much applies to quantum
fields as it does to random fields.
We could argue from this that quantum field models should as much be rejected as
random field models, but it seems more appropriate for physics to admit both.
The long-standing moratorium on construction of classical models loses most of
its justification if we allow ourselves to use the resources of random fields.

To temper the localism of this paper, repeating the caution given in section
\ref{Definitions}, a random field model that reproduces the phenomenological
success of a quantum field model has to have the same propagator as the
given quantum field model, which in classical terms is nonlocal even
while preserving signal locality and being Lorentz invariant \cite{Morgan}.
Thus the very first assumption that is needed to derive Bell inequalities for a
random field model, of locality in a classical dynamical sense, is not satisfied in
any realistic model, even though an assumption of signal locality is satisfied.
It is well known that quantum field theory is nonlocal in the sense of
Hegerfeldt \cite{Hegerfeldt}, while nonetheless preserving signal locality \cite{BY}.
The violation of Bell inequalities can be modelled by entirely local random fields,
but leaves an awkward question of how the nonlocal correlations might have been
established in the first place (that is, how did the ``conspiracy'' arise?),
which finds a relatively more natural answer if the propagator of a random
field model is nonlocal.

Although the principal argument of this paper is that correlations in random field
models generally do not satisfy the assumptions necessary to derive Bell inequalities,
empirically adequate random field models will often have to include a detailed
description of the measurement apparatus, which may well not be easy to construct.
If thermal properties of a measurement apparatus have to be taken into account explicitly
in a quantum field model to ensure empirical adequacy, however, then a random field model
should be no more complex than the quantum field model.
A random field approach might also connect better with general relativity because of their
shared classicality, but in my investigations so far the connection seems to be as awkward
as it is for quantum field theory.
Quantum theory is of course generally more easily usable than a random field model
whenever a finite dimensional Hilbert space is empirically adequate.

I am indebted to most of the people in and who have passed through Oxford's
Philosophy of Physics community, to Luca Porta Mana in Stockholm, and to
Stephen Adler in Princeton.

\appendix
\section{Beables}
\label{BeablesDescription}
The distinction Bell makes between ``beable'' fields and non-``beable'' fields, and the ontology that
Bell introduces, are not significant for the approach of this paper.
The only aspect of Bell's idea of ``beables'' that matters, both to his and to my mathematical derivation, is
the attachment of random variables to regions of space-time.
Nonetheless, Bell gives the name \textit{The theory of local beables} to the paper that is the principal stem of
the literature \cite[Chap. 7]{Bell}, so there should be a brief discussion here about beables.

As a first example, Bell distinguishes between the electromagnetic fields $\mathbf{E}$ and $\mathbf{H}$ as
``physical'' and the electromagnetic potentials $\phi$ and $\mathbf{A}$ as ``non-physical''; Bell emphasizes
that the connection is just a mathematical convenience that is ``not really supposed to \textit{be} there''
(from a random field perspective, the reason for discounting the 4-potential $A_\mu$ is more mathematical
than physical or ontological --- a classical connection $A_\mu$ cannot be averaged by integration over a
region, so it cannot be extended to a distribution that generates smeared observables, in contrast to
the electromagnetic field).
In the next paragraph Bell also denies that the wave function is a ``beable'' , for a somewhat different
reason, the ``$\,$`collapse of the wave function' on `measurement'$\,$'', which he describes as ``one of the
apparent non-localities of quantum mechanics''.
Bell resolves his ontological difficulties by claiming that the ``odd behaviour'' of the wave function is
acceptable if we take the wave function also to be only a mathematical convenience.
It's not quite clear what we should take the common feature of these examples to be, except perhaps the odd
behaviour (the electromagnetic potential is guilty only of ``funny behaviour''), which is the signal for
mathematics to be taken to be only a convenience instead of real.

Finally, he gives a description of what is important both for his and my purposes, ``We will be particularly
concerned with \textit{local} beables, those which (unlike for example the total energy) can be assigned
to some bounded space-time region'' (his emphasis).
It is manifest from the mathematics of section \ref{Derivation} that all that matters \textit{for the
mathematics} is the association of random variables with bounded regions of space-time.

\section{Continuous random fields}
\label{RandomFields}
For the purposes of this paper, continuous random fields can most appropriately be understood
either as random variable-valued distributions or, in a Koopman-von Neumann type of approach, as a
commutative quantum field (but see also \cite{Rozanov}).
We introduce either a random variable-valued linear map, $\chi:f\mapsto\chi_f$, or an
operator-valued linear map $\hat\chi:f\mapsto\hat\chi_f$, with the trivial commutator
$[\hat\chi_f,\hat\chi_g]=0$ whatever the space-time relationship between Schwartz space
functions $f$ and $g$ (a Schwartz space function $f(x)$ is infinitely often differentiable and
decreases as well as its derivatives faster than any power as $x$ moves to infinity in any
direction \cite[\S II.1.2]{Haag}).
The difference between these two approaches is mostly notational, but operator-valued
distributions are used in this appendix to emphasize the similarities to and differences
from quantum fields.
In contrast to the random field, for a quantized Klein-Gordon field, an operator-valued
linear map $\hat\phi:f\mapsto\hat\phi_f$, the commutator $[\hat\phi_f,\hat\phi_g]=(g,f)-(f,g)$
is zero when $f$ and $g$ have space-like separated supports.
$(g,f)$ is a manifestly Lorentz invariant Hermitian inner product on the Schwartz space,
\begin{eqnarray}
  (g,f)&=&\hbar\int \frac{\Intd^4 k}{(2\pi)^4}
                 2\pi\delta(k^\mu k_\mu - m^2) \theta(k_0) \tilde g^*(k) \tilde f(k)\\
       &=&\hbar\int \frac{\Intd^3 \mathbf{k}}{(2\pi)^3}
                 \frac{\tilde g^*(\mathbf{k}) \tilde f(\mathbf{k})}{2\sqrt{\mathbf{k}^2 + m^2}}.
\end{eqnarray}

Although we usually define the vacuum state of the quantized Klein-Gordon field in terms of the
trivial action of a creation operator, we can equally well define it by the characteristic function
\begin{equation}
  \left<0\right|e^{i\lambda\hat\phi_f}\left|0\right>=e^{-\Half\lambda^2(f,f)},
\end{equation}
which is enough to fix the Wightman functions of the quantum field (by linearity,
$e^{i\lambda\hat\phi_f+i\mu\hat\phi_g}=e^{i\hat\phi_{\lambda f+\mu g}}$, which we can use
to construct a multivariate characteristic function).
Other sectors can be constructed by changing the right-hand side, which, as well as thermal
sectors, include ``extra quantum fluctuation'' sectors \cite{MorganCKG},
\begin{equation}
  \left<0\right|e^{i\lambda\hat\phi_f}\left|0\right>=e^{-\Half\alpha\lambda^2(f,f)},\qquad\alpha>1
\end{equation}
($\alpha<1$ is not a state over the algebra of observables of the quantized Klein-Gordon field).
Analogously, we can define a state of the random field $\hat\chi_f$ by the characteristic function 
\begin{equation}
  \varphi_0(e^{i\lambda\hat\chi_f})=e^{-\Half\lambda^2(f,f)}.
\end{equation}
For this random field state, all joint probability densities over observables
$\hat\chi_{f_1},\hat\chi_{f_2},...,\hat\chi_{f_n}$ are identical to the equivalent joint probability
densities for the quantized Klein-Gordon field, whenever the quantum field observables are also
compatible.
Self-adjoint functions of non-commuting observables of the quantized Klein-Gordon field such as
$\hat\phi_f\hat\phi_g+\hat\phi_g\hat\phi_f$ will of course generally have different probability
densities from their random field equivalents.

The algebras of observables are not the same, but the state over the classical algebra is
sufficiently similar to the vacuum state over the quantum algebra to make it reasonable to call the
random field state a presentation of ``quantum fluctuations''.
Certainly the amplitude of the fluctuations of the random field is controlled by $\hbar$ and
the fluctuations are distinct from the thermal fluctuations of a classical Klein-Gordon field,
which can be presented as $\varphi_C(e^{i\lambda\hat\chi_f})=e^{-\Half\lambda^2(f,f)_C}$, with
the Lorentz non-invariant inner product
\begin{eqnarray}
  (g,f)_C&=&\kT\int\frac{\Intd^4 k}{(2\pi)^4}
                 \frac{2\pi\delta(k^\mu k_\mu - m^2) \theta(k_0)}{\Half k_0} \tilde g^*(k) \tilde f(k)\\
       &=&\kT\int \frac{\Intd^3 \mathbf{k}}{(2\pi)^3}
                 \frac{\tilde g^*(\mathbf{k}) \tilde f(\mathbf{k})}{(\mathbf{k}^2 + m^2)}.
\end{eqnarray}
This thermal state can be presented either with a trivial commutator $[\hat\chi_f,\hat\chi_g]=0$
or with the commutator $[\hat\chi_f,\hat\chi_g]=(g,f)_C-(f,g)_C$, depending on whether we wish to
use models in which idealized measurements are always compatible or generally not compatible at
time-like separation because of \emph{thermal} fluctuations (see \cite{MorganCKG}).

The difference between quantum fields and random fields can be taken to be only a different attitude
to idealized measurements; actual measurements can be described in terms of either.
The empirical principle that justifies the implicit description of quantum fluctuations that
underlies quantum theory is our apparent inability to reduce the quantum fluctuations of our
measurement apparatuses, in contrast to the almost universal minimization of thermal fluctuations
in precision experiments.
Even if this empirical principle is unbroken, however, we can still model quantum fluctuations and
their effects explicitly instead of implicitly, just as we model thermal fluctuations and
their effects explicitly when we have to.
It is perhaps a conceptual advantage that classical random fields explicitly describe thermal
and quantum fluctuations in the same way.
A mathematical model is valuable as a mental image of the world, not necessarily as how the
world really is; we can \emph{imagine} what the results of measurements might be if we had
classically ideal measurement devices, even if we don't have any.

More mathematics and discussion can be found in refs. \cite{Morgan} and \cite{MorganCKG}.

\section{Bell's original approach}\label{TraditionalBell}
The more general literature on Bell inequalities more-or-less follows Bell's
original approach \cite[Chap. 2]{Bell}, in that $\lambda$, $\mu$, and $\nu$ are
not distinguished by their space-time associations, but all hidden random variables
are instead written as a single set $\Lambda$.
Also, $a$ and $b$ are generally taken to be settings at the time of measurement, so that
they are associated with $\RA$ and $\RB$ instead of with $\PastRA - \PastRB$ and
$\PastRB - \PastRA$. Finally, $c$ is generally taken to be null.
However, only the lack of distinction between $\lambda$, $\mu$, and $\nu$ makes a
significant difference.
Following the analysis of Section \ref{Derivation}, the more general literature
(rationally reconstructed, since many different notations are used) writes
\VSbefore
\begin{eqnarray}
M(a,b) & = & \int\sum_{AB} AB\,p(A,B,\Lambda|a,b) d\Lambda \\
       & = & \int\sum_{AB} AB\,p(A,B|\Lambda,a,b)
                    p(\Lambda|a,b) d\Lambda \\
       & = & \int\sum_{AB} AB\,p(A|\Lambda,a,b)p(B|\Lambda,a,b)
                    p(\Lambda|a,b) d\Lambda \label{TBcompleteness}\\
       & = & \int\sum_{AB} AB\,p(A|\Lambda,a)p(B|\Lambda,b)
                    p(\Lambda|a,b) d\Lambda,\label{TBlocality}
\end{eqnarray}
in which the assumptions required to derive equations (\ref{TBcompleteness}) and
(\ref{TBlocality}) correspond to Jarrett's ``completeness'' and ``locality''
respectively \cite{Jarrett} (or ``outcome independence'' and ``parameter independence''
in Shimony's terminology \cite{Shimony}).
To allow Bell's original approach, it has to be assumed further that
$p(\Lambda|a,b)=p(\Lambda)$ (``no-conspiracy'').
If we take $\Lambda$ to be the microstate of a measured system, as we typically
do if we think we are measuring the state of two classical point particles,
correlation of $\Lambda$ with $a$ and $b$ represents contextuality of the
measured system state, which is generally taken to be anathema.
In a random field context, however, it is far more natural to take $\Lambda$ to be
the microstate of the whole experimental apparatus, because for general random fields
\emph{there is no natural way to draw an exact boundary} between what would usually
be termed the measurement device and the measured system, so that it seems that
$\Lambda$, when considered in the fullest possible detail, \emph{must} be correlated
with $a$ and $b$ (but this does not claim that $\Lambda$ causes or does not cause
$a$ and $b$).

The derivation of Bell inequalities given in Section \ref{Derivation} subsumes
the discussion that is possible if we do not distinguish $\lambda$, $\mu$,
and $\nu$, so we will not further pursue the limited approach of this Appendix.

\vspace{20pt}
\newcommand\JournalName[1]{\textit{#1}}
\newcommand\JournalVolume[1]{\textbf{#1}}
\newcommand\BookName[1]{\textit{#1}}
\newcommand\Preprint[1]{(\textit{Preprint} \texttt{#1})}

\end{document}